# Correct form of the electron wavefunction in periodic solids

T.R.S. Prasanna

Department of Metallurgical Engineering and Materials Science

Indian Institute of Technology, Bombay

Mumbai 400076 India

The Bloch wavefunction leads either to mathematically impossible consequences or suggests that the ground state energy is a function of size and shape when the geometry of large crystals is considered in detail. It is incompatible with the assumption underlying the Born-von Karman periodic boundary condition. The source of the difficulty is the incorrect dependence of the Bloch wavefunction on the wavenumber index $k$. The mathematically impossible consequences can be overcome if the periodic part of the electron wavefunction is represented as $u_n(\mathbf{r})$, which is dependent only on the band index, *n, and is independent of the wavenumber index $k$*. This correct form of the wavefunction is consistent with the Bloch theorem and with all other properties of Bloch wavefunctions. The correct form is also consistent with the Born-von Karman periodic boundary condition. The correct form of the electronic wavefunction in a periodic solid has profound consequences. It simplifies the calculation of electronic structure as only one wavefunction per band, $u_n(\mathbf{r})$, needs to be evaluated. It brings about a conceptual unification between the band picture favored by physicists and the bond picture favored by chemists. The correct form of the electron wavefunction will simplify the understanding of many phenomena involving valence electrons.

The Bloch wavefunction is the starting point for the application of Quantum Mechanics to periodic solids. The Bloch Theorem imposes conditions on the form of the electron wavefunction in periodic solids. This theorem states that due to the translational periodicity of the Bravais lattice (and the potential), the wavefunction of the electron has the same periodicity up to a phase factor and can be represented as

$$\psi_{nk}(r+R) = e^{ik.R} \psi_{nk}(r) \tag{1}$$

The Bloch wavefunction satisfies the Bloch Theorem and is given by

$$\psi_{nk}(r) = e^{ik.r} u_{nk}(r) \tag{2}$$

where $u_{nk}(r)$ is the periodic part and satisfies the relation $u_{nk}(r) = u_{nk}(r+R)$. The Bloch theorem is proved in different ways in standard textbooks [1-3]. Various other properties of the Bloch wavefunction are also described in standard textbooks [1,2].

In this paper, we show that the Bloch wavefunction, Eq.2, leads to mathematically impossible consequences or suggests that the ground state energy is a function of size and shape when the geometry of crystals is considered in detail. It is incompatible with the assumption underlying the Born-von Karman periodic boundary condition. The source of the difficulty is the incorrect dependence of the Bloch wavefunction on the wavenumber index *k*. The mathematically impossible consequences can be overcome if the periodic part of the wavefunction is represented as $u_n(r)$, which is dependent only on the band index, n, *and is independent of the wavenumber index k*. This correct form of the wavefunction is consistent with the Bloch theorem, Eq.1, and with all other properties of Bloch wavefunctions. This correct form of the electronic wavefunction in a periodic solid has profound consequences. It simplifies the calculation of electronic structure as only one wavefunction per band, $u_n(r)$, needs to be evaluated. It brings about a conceptual unification between the band picture favored by physicists and the bond picture favored by chemists. The correct form of the electron wavefunction will simplify the understanding of many phenomena involving valence electrons.

The Born-von Karman periodic boundary condition [1,3] is given by

$$\psi(r + N_i a_i) = \psi(r) \qquad (3)$$

The justification for the Born-von Karman periodic boundary condition is stated in Ref.1 as "*we adopt this boundary condition under the assumption that the bulk properties of the solid will not depend on the choice of boundary condition, which can therefore be dictated by analytical convenience*". Similar comments are found in Ref. 4 which states "*If the crystal is very large, we expect the precise form of these (boundary conditions) not to effect the physical description of properties over the bulk of the crystal. We may then choose (boundary) conditions which are most simple mathematically. These are the 'cyclic' or 'periodic' boundary conditions…*" Thus we see that the fundamental assumption behind the periodic boundary condition is that bulk properties of large crystals are unaffected by their size and shape. This assumption implies that the charge density within a unit cell inside a large crystal is unaffected by its size and shape or that the charge density within a unit cell is *exactly the same* in large crystals of different sizes and shapes.

Applying Bloch's Theorem, Eq.1, to the Born- von Karman periodic boundary condition, Eq.3, leads [1] to the allowed values of the wavenumber index, $k$. The wavenumber index is given by the relation $k = \kappa_1 b_1 + \kappa_2 b_2 + \kappa_3 b_3$ where $b_i$ are the reciprocal lattice vectors [3]. Also, $\kappa_i = n_i / N_i$ where $N_i$ are the total number of unit cells in each dimension and $n_i$ take integer values from 0 to $N_i$-1. This leads to the well known [1,3] result that the number of $k$ points in a Brilloiun Zone equals the total number of unit cells, N, in the crystal. The set of wavenumbers $\{k\}$ is uniquely determined by the size and shape [1,3] of a crystal. The electron is delocalized over the entire crystal. Therefore, the normalization condition of Bloch wavefunction, Eq.2, yields

$$\int_{uc} |\psi_{nk}(r)|^2 d^3r = \int_{uc} |u_{nk}(r)|^2 d^3r = 1/N \qquad (4)$$

as all N unit cells are identical. There are N periodic functions of the type $u_{nk}(r)$, one for each value of $k$, for a given band. Various spectroscopic experiments suggest that even in a solid different electron bands are distinguishable. Hence the total charge density can be written as the summation of charge densities in various bands. Under the assumption that bulk properties of large crystals are unaffected by their size and shape, the electron density for band $n$ is given by the sum

$$\rho_n(r) = \sum_k |u_{nk}(r)|^2 = \sum_{k'} |u_{nk'}(r)|^2 = \sum_{k''} |u_{nk''}(r)|^2 = \sum_{k'''} |u_{nk'''}(r)|^2 = \ldots \qquad (5)$$

Eq.5 represents the expression for charge density in crystals with different size, shape and number of unit cells, N, N′, N″ etc. In Eq.5, we have implicitly assumed that the band is fully occupied, but the arguments are independent of this assumption and are equally valid for partially filled bands. Each summation contains only positive terms. Eq.5 contains an infinite number of equalities, one for each size and shape of a crystal. The Bloch wavefunction, Eq.2, implicitly assumes that the periodic part of the wavefunction, $u_{nk}(r)$, is different and distinct for different values of wavenumber, *k*. Otherwise, there would be no need for the wavenumber index, *k*, to be part of the description of $u_{nk}(r)$. This dependence of $u_{nk}(r)$ on the wavenumber index, *k*, leads to mathematically impossible consequences as discussed below.

We consider (without loss of generality) cubic crystals whose dimensions are more than 1-2 *μ*m in each dimension to avoid size effects prevalent in nanometer size crystals. An average lattice parameter is of the order of 3-4 A, and hence, we consider a minimum number of 4000 or $2^{12}$ unit cells along each dimension in a crystal. Therefore, a three dimensional crystal is considered to have a minimum number of 64 $10^9$ or $2^{36}$ unit cells. However, there is no theoretical upper limit on the number of unit cells in a crystal. Many single crystals, e.g. Si, are routinely grown to sizes of inches and feet. Here, a reasonable upper limit for the size of a crystal is considered to be higher by approximately four orders of magnitude in each dimension than the lower limit, which would make each dimension to be of the order of a few centimeters. Therefore, in such a large crystal 64 $10^{21}$ or $2^{75}$ unit cells would be present. We consider the consequences of Eq.5 on different crystals below.

**Case A - *Two crystals of identical shape but different sizes*.** Consider 2 large cubic crystals ($C_1$ and $C_2$) with dimensions in the centimeter range or $2^{25}$ unit cells in each dimension. If in one of these crystals, $C_2$, each dimension is repeatedly halved and new crystals labeled $C_{2'}$, $C_{2''}$, etc. then the number of unit cells is reduced by $\frac{1}{2}^3$ for each halving. Since the number of unit cells in each dimension, $N_i$, is reduced, the number of $k$ points is also reduced appropriately. *However, it is important to note that no new $k$ points are generated, i.e. only 1 of every 8 $k$ points is retained for each halving.* According to Eq.5, the charge density, $\rho_n(r)$, is *exactly the same* in both crystals ($C_1$ and $C_{2'}$) even though the summation in Eq.5 is over different number of terms. The summation for crystal $C_1$ contains all the same terms as crystal $C_{2'}$ but contains 7 times additional terms that are all different yet both summations give *exactly the same* charge density, adjusted for normalization. This process can be continued and the second crystal can be halved till each dimension is $2^{12}$ unit cells (in the $\mu$m range) and this crystal can be labeled $C_{2'''}$. Comparison of crystals $C_1$ and $C_{2'''}$ show that *exactly the same* charge density must be obtained by the summation given by Eq.5. Again, the $k$ points in $C_{2'''}$ are a subset of those in crystal $C_1$ and no new $k$ points are introduced. According to Eq.5, the charge density for $C_{2'''}$ contains a summation over $2^{36}$ different terms whereas for $C_1$ contains a summation over $2^{75}$ different terms of which $2^{36}$ are the same terms as in $C_{2'''}$. Considering that each summation contains only positive non-zero terms, it is impossible to conceive of any type of mathematical functions that can give *exactly the same* charge density (adjusted for normalization) when astronomical numbers of different functions are summed over their magnitude squared.

**Case B - *Two crystals of identical size but different shapes*.** Consider two crystals, $C_3$ and $C_4$, with the same number of unit cells, N, but different shapes due to which the set of wavenumbers $\{k\}$ will be different for each crystal. As in Eq.5, the charge density is expressed as $\rho_n(r) = \sum_k |u_{nk}(r)|^2 = \sum_{k'} |u_{nk'}(r)|^2$. While the number of terms in the summation are identical (N), the functions $u_{nk}(r)$ and $u_{nk'}(r)$ will be different as the set of wavenumbers $\{k\}$ is different due to the different shapes of the two crystals. Therefore, the charge density is given by a summation over an identical number of terms, N, in both crystals and is *exactly the same* even though each term, $|u_{nk}(r)|^2$, in the summation for crystal $C_3$ is different from each term, $|u_{nk'}(r)|^2$, in the summation for crystal $C_4$. Considering that each summation contains $2^{36}$ or $10^{10}$ or more terms, this requirement is clearly a mathematical impossibility.

**Case C - *Multiple crystals of different sizes and shapes*.** A large crystal in broken into powder form generating new crystals of arbitrary size and shape in the range of 1-1000 microns. It is easy to obtain millions of such crystals in practice and for the purposes of this argument we consider $10^6$ crystals with random shapes in the 1-1000 microns range. Because of the random shapes, each crystal will have different set of wavenumbers $\{k\}$ and hence, different functions $u_{nk}(r)$. The charge density is *exactly the same* in all crystals, i.e. in $10^6$ summations of the type Eq.5, one for each crystal. Each summation contains $10^{10} - 10^{19}$ different terms, $|u_{nk}(r)|^2$, and all (~ $10^6$) summations give *exactly the same* charge density, adjusted for normalization. The above requirement is clearly a mathematical impossibility.

From the above discussion, we see that Eq.5 leads to mathematically impossible consequences. Because the periodic part of the Bloch wavefunction, $u_{nk}(r)$, is different for different values of *k* and the set of wavenumbers {*k*} is uniquely determined by the size and shape of crystals, the band charge density, $\rho_n(r)$, will depend on size and shape and cannot be exactly the same in large crystals. *Hence, the Bloch wavefunction does not satisfy Eq.5*. Summing over band charge densities, the total charge density, $\rho(r)$, of large crystals will depend on their size and shape. It follows from the Density Functional Theorem [5] that the ground state energy of large crystals will depend on their size and shape. Therefore, it follows that all bulk properties of large crystals will depend on their size and shape, which is exactly opposite to the fundamental assumption underlying Born-von Karman periodic boundary conditions. *Therefore, the Bloch wavefunction is incompatible with Born-von Karman periodic boundary condition*. Clearly both of them cannot be correct and one of them is in error. The assumption underlying Born-von Karman periodic boundary condition, *viz*. bulk properties of large crystals are independent of size and shape, is unlikely to be in error as it is difficult to conclude that the charge density within a unit cell deep inside a large crystal will be affected by its size and shape. If it were true, it would suggest that ground state energy and bulk properties of large crystals must be specified with reference to their size and shape. This is not supported by any experimental observations of bulk properties. Hence, any error in the assumption underlying Born-von Karman periodic boundary condition can be ruled out.

Therefore, the likely source of error is in the Bloch wavefunction. We see from applying Eq.5 to cases A, B and C above that the Bloch wavefunction, Eq.2, leads to mathematically impossible consequences. Since $u_{nk}(r)$ depends on only two indices, *n* and *k*, and dependence on *n* is universal in all quantum systems, it is clear that the dependence of the wavefunction on the index *k* is the source of the problem. *Hence, the Bloch wavefunction, Eq.2, contains an incorrect dependence on the index **k**.*

A resolution to the difficulty can be found if the periodic part of the wavefunction, $u_{nk}(r)$, is *independent of **k***. Therefore, it is postulated that the form of the electronic wavefunction in periodic solids is given by

$$\psi_{nk}(r) = e^{ik\cdot r}\, u_n(r) \qquad (6)$$

That is, the periodic part of the wavefunction is identical for all values of wavenumber *k*. By simple substitution it is readily seen that this wavefunction satisfies the Bloch Theorem, Eq.1. It also satisfies all other properties of electronic wavefunction in periodic solids, i.e. properties required of Bloch wavefunction, Eq.2, discussed in standard textbooks [1,2]. This is only to be expected as the correct electronic wavefunction, Eq.6, is a special case of the Bloch wavefunction, Eq.2, where all $u_{nk}(r) = u_n(r)$ and will therefore satisfy all conditions that Bloch wavefunction satisfies. This form of the electron wavefunction readily overcomes the conditions (Cases – A, B and C) imposed by the geometry of crystals described above. This is because each term in every summation in Eq.5 is identical, i.e. $|u_n(r)|^2$, and changing the shape and size of crystal does not alter this function but only alters the normalization constant. Hence, Eq.6 provides a simple explanation for cases A, B and C, and more generally Eq.5, where different (infinite) summations, each containing astronomical number of terms, give *exactly the same* charge density. *The correct form of the wavefunction, Eq.6, is* *compatible with the Born-von Karman periodic boundary condition.* Therefore, Eq.6 can be considered to be the correct form of the electron wavefunction in a periodic solid.

Because the Bloch wavefunction has been in use for 75 years without attention being paid to the consequences of geometry of crystals, it is necessary to recapitulate the *theoretical difficulties* posed by it. The Bloch wavefunction leads to requirements that are mathematically impossible when the geometry of crystals in considered in detail. Therefore, at least a plausible justification if not a rigorous mathematical proof for the existence of mathematical functions that would satisfy cases A, B and C discussed above is necessary *without which the use of Bloch wavefunction is theoretically unjustified*.

The correct form of the electronic wavefunction, Eq.6, has profound consequences. When Eq.6 is substituted for the electron wavefunction, the Schrodinger's equation for periodic solids is modified to

$$\left[\frac{-\hbar^2}{2m}(\nabla + i\mathbf{k})^2 + V(\mathbf{r})\right] u_n(\mathbf{r}) = \varepsilon_{n\mathbf{k}} u_n(\mathbf{r}) \qquad (7)$$

where $\varepsilon_{n\mathbf{k}}$ is the energy level of the electron wavefunction, $\psi_{n\mathbf{k}}(\mathbf{r})$, given by Eq.6. The first point to note is that it is only necessary to solve for one function per band, $u_n(\mathbf{r})$. Once $u_n(\mathbf{r})$ is known, determining $\varepsilon_{n\mathbf{k}}$ for all values of the wavenumber $\mathbf{k}$ is a simple matter of substituting different values of $\mathbf{k}$ in Eq.7 and performing appropriate computations. However, it is important to note that $u_n(\mathbf{r})$ is not necessarily the function that is obtained by solving Eq.7 for $\mathbf{k} = \mathbf{0}$, i.e. that function that leads to the lowest energy for $\varepsilon_{n0}$ at $k = 0$. The function $u_n(r)$ is one that minimizes the total band structure energy, i.e. the band energies $\varepsilon_{nk}$ summed over all occupied $k$ states.

The kinetic energy is obtained from Eq.7 as

$$KE = \frac{-\hbar^2}{2m}\int_{uc} u_n^*(r)\nabla^2 u_n(r)d^3r - \frac{\hbar^2 2ik}{2m}\int_{uc} u_n^*(r)\nabla u_n(r)d^3r + \frac{\hbar^2 k^2}{2m} \qquad (8)$$

As $k$ changes, there is no change in the first term which *has to be evaluated only once* for each band. The second term involves determination of an integral that *has to be evaluated only once* for each band and depends linearly on $k$. If the Bloch wavefunction, Eq.2, were used instead, it would result in functions $u_{nk}(r)$ in the integrals in the first two terms, which then would have to be determined for each value of $k$. Hence, the correct form of the electronic wavefunction results in a significant reduction of computational effort.

The potential energy is given by PE = $E_{e-e} + E_{n-e} + E_{xc} + E_{n-n}$ where all the terms have their usual meaning. The nuclear-nuclear repulsion term, $E_{n-n}$, is usually evaluated by the Ewald summation technique [3] that uses artificial parameters. Recently, we have described the correct method [6] to evaluate this term by incorporating zero point vibrations. The electron-electron repulsion energy term between <u>two electrons</u> is given by

$$E_{e-e} = \iint \frac{|u_n(r_i)|^2 |u_{n'}(r_j)|^2}{|r_i - r_j|} d^3r_i\, d^3r_j \qquad (9)$$

<u>This shows that the electron-electron repulsion energy between two electrons is independent of $k$.</u> It is readily seen that the nuclear-electron repulsion energy is also independent of $k$. Therefore, it is necessary to calculate $E_{e-e}$ and $E_{n-e}$ only once for each combination of bands, $n$ and $n'$, unlike in the Bloch form of the wavefunction, Eq.2, where they have to be calculated for each value of $k$ separately. This also leads to an enormous reduction in computational effort.

The exchange energy between <u>all electrons in bands $n$ and $n'$,</u> when the Bloch form, Eq.2, is used is given by

$$E_{ex}^{Bl} = \sum_{k}\sum_{k'} \iint \frac{u_{nk}^*(r_i)\, u_{n'k'}^*(r_j)\, u_{nk}(r_j)\, u_{n'k'}(r_i)}{|r_i - r_j|}\, e^{i(k'-k)\cdot(r_i - r_j)}\, d^3 r_i\, d^3 r_j \qquad (10)$$

This term is simplified in the correct form of the electronic wavefunction, Eq.6, to

$$E_{ex}^{new} = \sum_{k}\sum_{k'} \iint \frac{u_{n}^*(r_i)\, u_{n'}^*(r_j)\, u_{n}(r_j)\, u_{n'}(r_i)}{|r_i - r_j|}\, e^{i(k'-k)\cdot(r_i - r_j)}\, d^3 r_i\, d^3 r_j \qquad (11)$$

If some suitable assumptions are made regarding the shape of the Fermi Surface, the double summation over $k$ and $k'$ can be replaced by an integral and after integrating the phase factor over variables $k$ and $k'$ and calling the resulting function $f(r_i, r_j)$, the exchange integral between <u>all electrons in bands $n$ and $n'$</u> can be rewritten as

$$E_{ex}^{new} = \iint \frac{u_{n}^*(r_i)\, u_{n'}^*(r_j)\, u_{n}(r_j)\, u_{n'}(r_i)}{|r_i - r_j|}\, f(r_i, r_j)\, d^3 r_i\, d^3 r_j \qquad (12)$$

Eq.12 shows that the expression for exchange energy in periodic solids between all electrons in bands $n$ and $n'$ is very similar to that in atoms and molecules (where $n$ and $n'$ would represent energy levels) but differs in that the integral contains an additional function $f(r_i, r_j)$. This additional function is due to the presence of many electrons in a given band. Therefore, evaluating the exchange energy in the local density approximation (LDA) using the same approximation ($E_{ex} \propto \rho(r)^{1/3}$) for both molecules and solids [7,8] is not appropriate. Another interesting possibility is that since for some molecules with small number of electrons it is possible to evaluate the exchange energy integral, Eq.12 suggests that the same may be possible for some solids as well.

In the correct form of the electronic wavefunction, Eq.6, the Schrodinger's equation in periodic solids is solved for functions, $u_n(r)$, that minimize the total energy. The bond picture, $|u_n(r)|^2$, is directly obtained from $u_n(r)$. The band picture, $\varepsilon_{nk}$ vs $k$, can be obtained once $u_n(r)$ is known. *Hence, both the band and bond pictures are central to electronic structure calculation.* The goal of electronic structure calculations can be described as to obtain the correct $u_n(r)$ (bond picture) so as to minimize the total energy (band picture). This is philosophically identical to the approach adopted in calculating electronic structure of atoms and molecules where the Schrodinger's equation is solved for the correct wavefunctions, $\psi_n(r)$, (bond picture) that give the lowest total energy (equivalent of the band picture). Therefore, while the details of various terms are different, philosophically, the solution of the Schrodinger's equation for periodic solids and molecules are similar in spirit, which reconciles the *band* and *bond* pictures, i.e. the physicists and chemists approach to electronic structure. This is unlike the present state where the Schrodinger's equation is solved to obtain the Bloch wavefunctions, Eq.2, which give the lowest total energy and only the band picture is important. The bond picture is obtained afterwards by Eq.5 and is not central to electronic structure calculations.

The electron wavefunction is the starting point for explaining many phenomena involving valence electrons in solids. Since the correct form of the electron wavefunction, Eq.6, is simpler than the Bloch wavefunction, Eq.2, it follows that it would simplify the understanding of many phenomena involving valence electrons. For example, for low lying excitations in metals where electrons occupy new states in the same band, to first order approximation, the periodic part of the wavefunction, $u_n(r)$, remains unchanged while only the phase factor $e^{ik \cdot r}$ will be altered. Therefore, to first order, the coulomb interaction energy terms remain *unchanged* as seen from Eq.9. The change in electron energy on excitation is dominated by the kinetic energy terms with a (small) contribution from the exchange energy term. In other words, the additional energy acquired by the electron upon excitation is mostly in the form of kinetic energy. Clearly, there will be many phenomena involving valence electrons whose explanations will be simplified when the correct form of the wavefunction, Eq.6, is used.

In conclusion, the Bloch wavefunction leads either to mathematically impossible consequences or suggests that the ground state energy is a function of size and shape when the geometry of large crystals is considered in detail. It is incompatible with the assumption underlying the Born-von Karman periodic boundary condition. The source of the difficulty is the incorrect dependence of the Bloch wavefunction on the wavenumber index $k$. It is proposed that the correct form of the wavefunction of an electron in a periodic solid is $\psi_{nk}(r) = e^{ik \cdot r} u_n(r)$, i.e. the periodic part, $u_n(r)$, is dependent only on the band index, $n$, and is independent of the wavenumber index $k$. This correct form of the wavefunction is consistent with the Bloch theorem and with all other properties of Bloch wavefunctions. The correct form is also consistent with the Born-von Karman periodic boundary condition. The correct form of the electronic wavefunction in a periodic solid has profound consequences. It simplifies the calculation of electronic structure as only one wavefunction per band, $u_n(\mathbf{r})$, needs to be evaluated. It brings about a conceptual unification between the band picture favored by physicists and the bond picture favored by chemists. The correct form of the electron wavefunction will simplify the understanding of many phenomena involving valence electrons.